\documentclass[aip,apl,reprint,graphicx]{revtex4-1}

\usepackage{graphicx}% Include figure files
\usepackage{dcolumn}% Align table columns on decimal point
\usepackage{bm}% bold math
\usepackage{amsmath}
\usepackage{color}
\usepackage{textcomp}

\usepackage{hyperref}

\begin{document}

\preprint{}

\title{Single shot all optical switching of intrinsic micron size magnetic domains \\ of a Pt/Co/Pt ferromagnetic stack}

\affiliation{Universit\'e de Strasbourg and CNRS, IPCMS UMR 7504, Strasbourg, France}%

\author{M. Vomir}%
%\affiliation{Universit\'e de Strasbourg and CNRS, IPCMS UMR 7504, Strasbourg, France}%
\email{vomir@ipcms.unistra.fr}

\author{M. Albrecht}%
%\affiliation{Universit\'e de Strasbourg and CNRS, IPCMS UMR 7504, Strasbourg, France}%

\author{J.-Y. Bigot}%
%\affiliation{Universit\'e de Strasbourg and CNRS, IPCMS UMR 7504, Strasbourg, France}%
\email{bigot@ipcms.unistra.fr}

\date{\today}

\begin{abstract}
We demonstrate that the magnetization reversal in a ferromagnetic Pt/Co/Pt stack can be induced by a single femtosecond laser pulse.  We find that the size of the switched spot is comparable to the size of the intrinsic magnetic domains. It requires an absorbed energy density of $\sim$4~mJ.cm$^{-2}$, beyond which the excited spot fragments into a multidomain structure. The switching process occurs back and forth with subsequent laser pulses and it is helicity-independent.  Furthermore, the dynamics of the magnetization reversal occurs in a timescale less than one microsecond. These results suggest that all optical switching in ferromagnetic films requires to match the laser spot with the specific domain sizes.
\end{abstract}


\maketitle %\maketitle must follow title, authors, abstract and \pacs


The all optical switching (AOS) of magnetic materials, first demonstrated in ferrimagnetic GdFeCo thin films \cite{Stanciu2007} has recently been extended to ferromagnetic Co/Pt thin films and FePt nanoparticles\cite{Lambert2014}. In the case of GdFeCo the mechanism of the magnetization reversal is due to the different dynamics of the two exchanged coupled sublatices \cite{Radu2011}. The corresponding switching time is of the order of tens of picoseconds \cite{GorchonPRB2016}. Alternatively, an electronic heat current induced and propagating in a thick Pt/Au layer deposited on top of GdFeCo results in faster switching times\cite{WilsonPRB2017}.  The reported switching in the ferromagnetic materials is helicity dependent and results from a cumulative  multi-pulse process\cite{Lambert2014,ElHadri2016,Tsema2016,Medapalli2016,Takahashi2016,John2017}. Indeed, as previously proposed\cite{Cornelissen2016,Gorchon2016}, the first laser pulse spontaneously induces a multi domain structure that further expands upon successive pulses. The Inverse Faraday Effect (IFE) has also been considered as the driving mechanism for the cumulative switching process of ferromagnets \cite{Cornelissen2016}. In addition, exchange coupled ferro(Co/Pt)-ferri magnetic stacks have been shown to lead to a picosecond switching under helicity independent excitation \cite{GorchonAPL2017}.

Here we show for the first time that the magnetization of a single ferromagnetic Co/Pt stack can be reversed by a single femtosecond laser pulse. In addition, the switching is a reversible process that occurs with linearly polarized pump pulses. The size of the switched spots is shown to be of the order of the intrinsic static domains. Furthermore, by combining several time resolved complementary techniques to resolve the dynamics of the magnetization reversal we show that the switching induced by femtosecond individual pulses occurs in a timescale of less than a few microseconds.

\begin{figure}
\begin{center}
\includegraphics[scale=0.39]{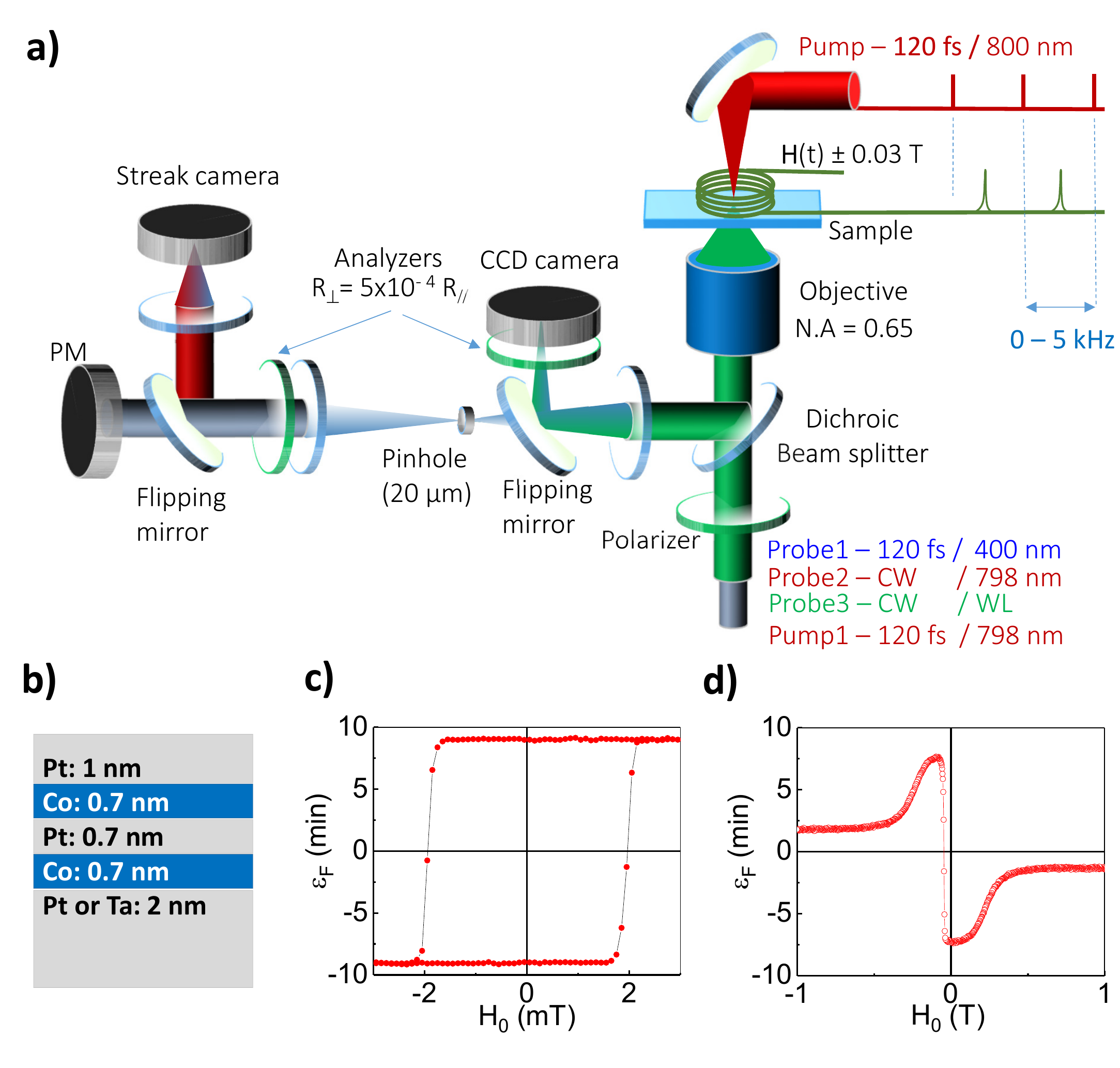}
\caption{\label{fig1} Experimental configurations of the confocal magneto-optical Kerr microscope (a). Sketch of the samples structure (b). Static magnetization loops measured perpendicular (c) and parallel (d) to the sample plane.}
\end{center}
\end{figure}

The experimental configurations of the magneto-optical confocal microscope used for our experiments is sketched in Fig.~\ref{fig1}(a). The pump pulses of 120~fs at a central wavelength of 800 nm have a repetition rate of 1~kHz. A synchronous mechanical chopper decreases the repetition rate of the laser to 10Hz, while a synchronous shutter allows, when required, single shot excitations of the sample. The pump pulses are focused either on the back side of the sample through the glass substrate within a spotsize of 30 \textmu m, or on the front side of the sample within a spotsize of 0.8 \textmu m using a 0.65 numerical aperture objective.
Several configurations are available for the probe. First, the magnetization dynamics with high temporal resolution (100~fs to 1~ns) is measured with femtosecond probe pulses (frequency doubled to 400~nm) focused on the front side of the sample trough the objective lens and collected back with a dichroic beam splitter \cite{Laraoui2007}. Second, for longer time delays (hundreds of~ns to tens of~\textmu s) we use a continuous laser diode in combination with a spectrometer and streak camera detection. Finally, for static analysis of the magnetization switching we image the sample plane on a CCD camera using a spectrally filtered tungsten lamp.
In all configurations the polarization of the probe beam is further analyzed using the crossed polarizers technique with an extinction ratio of 5x10$^{-4}$. For the stroboscopic experiments, a pulsed magnetic field of 50~\textmu s and up to 0.03~T resets the magnetization in between each laser pulse.
Two sets of samples were sputtered on glass substrates either with or without a thicker (2~nm) buffer of Pt or Ta. The samples without buffer are symmetric sandwiches of Pt$_\text{x}$/Co$_\text{0.7nm}$/Pt$_\text{x}$ with x = 0.7; 1; 1.5; 2; 3 nm. Typical static hysteresis loops measured parallel and perpendicular to the film plane are shown in Fig.~\ref{fig1}(c) and (d) for the two stack sample: glass/Ta$_\text{2nm}$/[Co$_\text{0.7nm}$/Pt$_\text{0.7nm}$]x2/Pt$_\text{0.3nm}$.

\begin{figure}
\begin{center}
\includegraphics[scale=0.54]{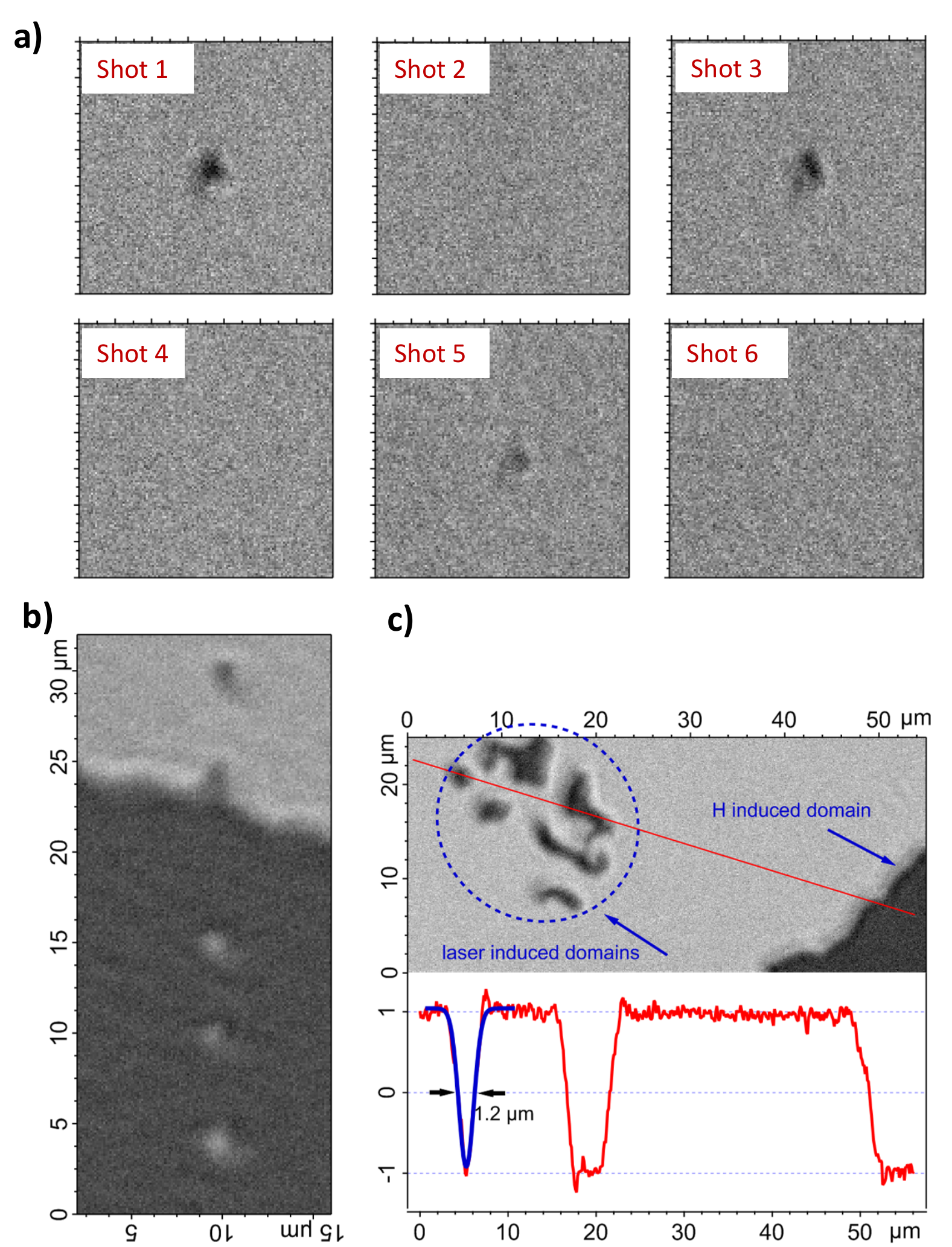}
\caption{\label{fig2} Magnetization switching for consecutive single pump laser pulses (a). Single shot AOS with linearly polarized laser pulses toggling between the two opposite magnetization directions (b). Switching across large static domains prepared with the magnetic field (c). The projection of a cross section along the red line is displayed in the bottom panel with a Gaussian fit (FWHM of 1.2~\textmu m) of a small size magnetic domain.}
\end{center}
\end{figure}

The magnetization reversal using a single laser pulse focused to 0.8~\textmu m is displayed in Fig.~\ref{fig2}(a) for 6 consecutive pulses. The sample used here is a single stack of Pt$_\text{1.5nm}$/Co$_\text{0.7nm}$/Pt$_\text{1.5nm}$. It shows that the linearly polarized laser pulses (LP) toggle the magnetic domain between the two opposite perpendicular directions. Similar results (not shown here) are obtained using left (LP) or right circular (RP) polarized pulses, demonstrating that the reversal process is polarization insensitive. Furthermore, as shown in Fig.~\ref{fig2}(b) for a prepared domain structure containing large magnetic domains, the LP pulses also switch the magnetization to the opposite state, indicating that the reversal process of a single magnetic domain is independent on the polarization. The density of the laser pulse for which the single domain reversal occurs  is E$_0$~=~4~mJ.cm${-2}$. This value is within a sharp window ($\sim$3~\%) between under which no reversal occurs and beyond which breaking into  a multidomain structure as the one shown in Fig.~\ref{fig2}(c). It is obtained for a larger laser pump spot, exciting from the back side of the sample.

A comparison between the magnetic domains obtained by magnetic field sweep and the laser excitation above the threshold and within a larger spot area is shown in Fig.~\ref{fig2}(c) for the same sample Pt$_\text{1nm}$/Co$_\text{0.7nm}$/Pt$_\text{1nm}$. A cross section of the image along the red line is projected in the bottom panel of Fig.~\ref{fig2}(c). It shows several magnetic domain sizes with smallest diameter of $\sim$1.2~\textmu m (fully resolved with our imaging system) as fitted with a Gaussian function.
\begin{figure}
\begin{center}
\includegraphics[scale=0.55]{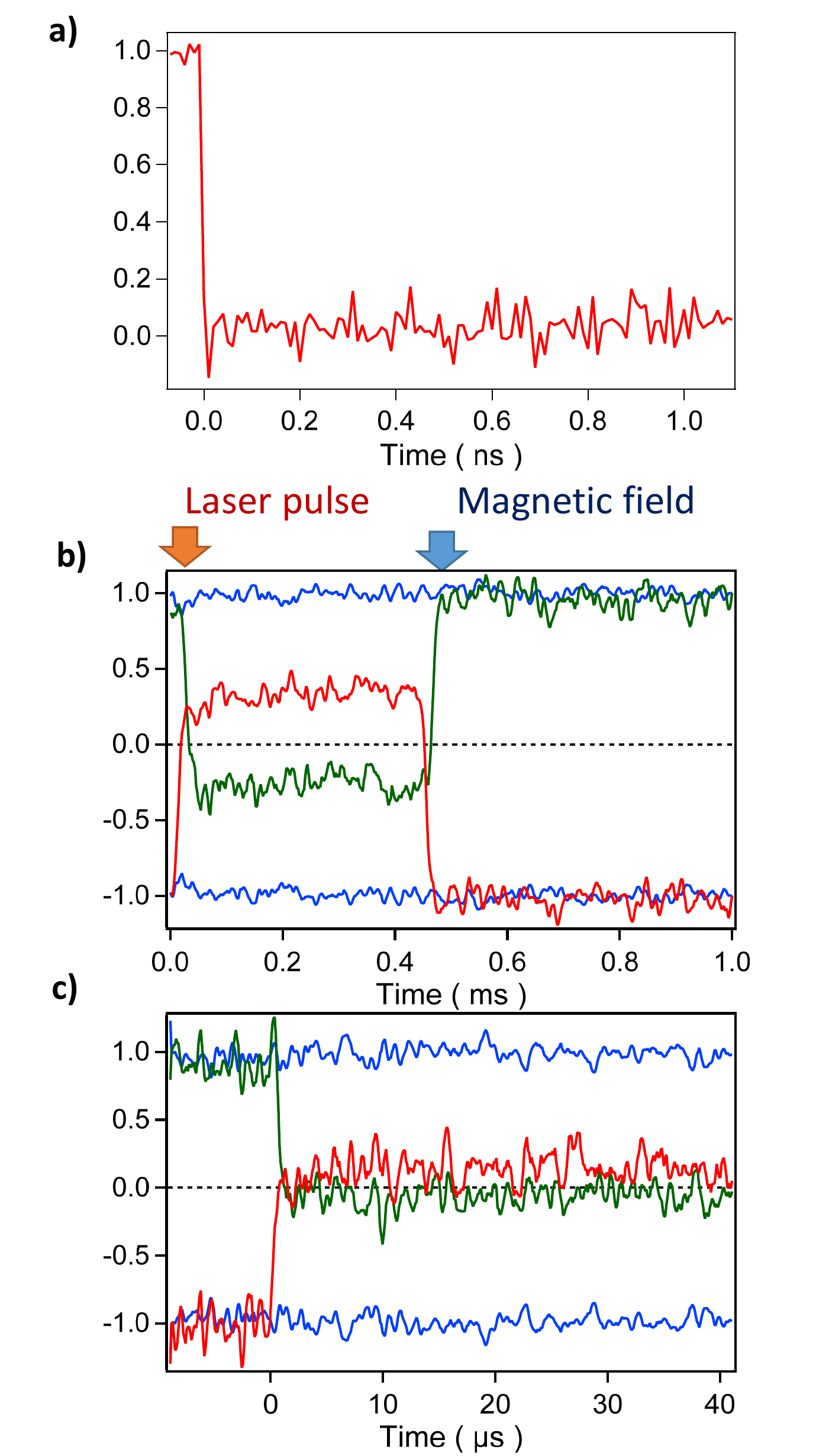}
\caption{\label{fig3} Time resolved magnetization dynamics. During the first nanosecond using the femtosecond pump-probe configuration (a), in the temporal windows of 1~ms (b), and 40~µs (c) using the streak camera. (see text for the details of each curve)}
\end{center}
\end{figure}
The magnetization dynamics during the first nanosecond is shown in Fig.~\ref{fig3}(a) using the configuration of the microscope with the pump from the substrate side.  After the ultrafast full demagnetization, characteristic of ferromagnetic films, the sample remains demagnetized, without any indication of reversal (zero crossing). As expected, the initial step occurs within a few hundreds femtoseconds (not shown here). To resolve the switching dynamics at longer time delays, the configuration with the streak camera is used. The results obtained for two temporal windows are presented in Fig.~\ref{fig2}(b)-(c) zooming from 1~ms to 40~\textmu s full scale. In each figure the initial up or down states are represented by the blue lines. The red (green) line corresponds to the reversal from the upper (lower) state respectively. Defining the switching time as the zero crossing point of the signals, Fig.~\ref{fig2}(c) indicates that it is less than a few microseconds. It is of the order of the temporal resolution for the particular voltage of the streak camera. Even though it has a 30 ps time resolution for the largest voltage ramp, we are limited here by the reduced number of photons present after the analyser in the magneto-optical microscopy configuration.

The above results suggest that the switching occurs via a thermal process when the spot size of the pump is of the order of the intrinsic domains size d. It is well known that, in ferromagnetic materials, d varies with the parameters of the sample such as its thickness T, the anisotropy constant K$_u$ (perpendicular in the present case) and the magnetization at saturation M$_s$ \cite{Malek1958,Kooy1960}. In addition, due to the time dependent temperature, d is also a function of time. To verify our assumption of matching the beam spot size with the domains we have modelled the dimensions of the static domains obtained after switching with a larger pump spot diameter (Fig.~\ref{fig2}(c)). The complexity of the problem requires crude approximations. Following Z. Malek et al. \cite{Malek1958}, we consider a domain structure made of periodic adjacent stripes in an infinite plane (the one of the sample), with alternative up and down perpendicular magnetic domains. The domain size d is obtained by minimizing, with respect to d, the total energy E resulting from the demagnetization energy and the energy of the domain walls, which dimensions are considered much smaller than the magnetic domains.
\begin{eqnarray}
E=\frac{16M^{2}_s}{\pi^2}d\sum^{\infty}_{n=1,n~\text{odd}}\frac{1}{n^3}
    \left(1-e^{-n\frac{\pi T}{d}}\right)+\frac{\gamma T}{d} ;\nonumber  \\ \gamma=4\sqrt{AK_u}
\end{eqnarray}
A is the exchange interaction parameter. The minimization procedure therefore leads to find the roots of the transcendental equation:
\begin{eqnarray}
\sum^{\infty}_{n=1,n~\text{odd}}\frac{1}{n^3}
    \left[1-e^{-nX}(1-nX)\right]=\widetilde{\gamma}X^2; \nonumber  \\
    \text{with}~\widetilde{\gamma}=\frac{\gamma}{16TM^2_s}; X=\frac{T}{d}
\end{eqnarray}
Using the experimental values: K$_u$=10$^7$~erg.cm$^{-3}$, M$_s$=2900~emu.cm$^{-3}$, T=0.7$\cdot10^{-7}$~cm, A=1.2$\cdot10^{-6}$~erg.cm$^{-1}$ we obtain $d$=0.56$\cdot10^{-4}$~cm. This value is twice smaller than the observed one (Fig.~\ref{fig2}(c)), which is satisfying given the crude approximations made. The anisotropy constant K$_u$ is obtained by fitting the asymptotic behavior of the hysteresis curve of the Pt$_\text{1.5nm}$/Co$_\text{0.7nm}$/Pt$_\text{1.5nm}$ sample following the model of H. Zhang et al. \cite{Zhang2010}. Let us emphasize that the above model, also reported in similar Co/Pt multilayers \cite{ElHadriPRB2016}, just gives an estimate of the domain size as it does not consider a random distribution of domains nor the finite size of the domain structure which is not infinite but mostly limited by the pump beam diameter.

Regarding the switching dynamics, one has to distinguish two different temporal regimes. At ultrashort times (Fig.~\ref{fig2}(a)), the demagnetization takes place in a few hundreds of femtoseconds and M(t) remains 0 during the observed window of 1~ns. This behavior can be described as in many ferromagnetic materials by a three temperature model for the charges, the spins and the lattice taking into account the time dependent anisotropy \cite{Beaurepaire1996,JYB2005,}. The very initial demagnetization is a non-thermal process which is due to a combination of spin-orbit and diffusive spins \cite{Shokeen2017}. The longer temporal window (Fig.~\ref{fig2}(b) and (c)), is associated to the switching time $\tau_{switch} \leq$ 1 \textmu s. The associated mechanism is unclear at the present time but most likely occurs via a nucleation process. Before the magnetization reaches an equilibrium multi-domain structure as the one displayed in Fig.~\ref{fig2}(c), starting from the fully demagnetized situation in Fig.~\ref{fig3}(a), nucleation spots appear and evolve in time. When the size of the "intrinsic" domains is comparable to the heated spot (Fig.~\ref{fig2}(a)), which occurs for a particular pump energy density, a single reversed domain is formed. As the process is helicity independent, a description of the toggling from pulse to pulse between the up and down magnetic states has to include the boundary conditions at the perimeter of the laser spot or heat spot (depending on the transverse thermal diffusion). That can be deduced from the static domain structure obtained for large pump energy densities, well above the one required to obtain a single switched domain (not shown here). It consists of an outer switched perimeter and an inner random distribution of domains\cite{Laraoui2007PhD}. Further theoretical investigations would require a dynamical model of the thermally induced nucleation/propagation process with the constraint of a static magnetization surrounding the laser spot (or thermal spot).

In conclusion, we have shown that all optical magnetization reversal in Co/Pt multilayers can be triggered by a single femtosecond pulse. The process is reversible for consecutive pulses and does not depend on the polarization of the pump. The novelty of the magneto-optical time resolved technique used in our studies demonstrates that the AOS switching in ferromagnetic multilayers is a relatively slow process occurring in the microsecond time scale.

\begin{acknowledgments}
The authors thank G. Versini and M. Acosta, for the samples preparation. We acknowledge support from the Agence Nationale de la Recherche in France via the projects EquipEx UNION No. ANR-10-EQPX-52.

\end{acknowledgments}

% If in two-column mode, this environment will change to single-column format so that long equations can be displayed.
%\begin{widetext}
%$$\mbox{put long equation here}$$
%\end{widetext}

% Figures should be put into the text as floats.
% Use the graphics or graphicx packages (distributed with LaTeX2e).
% Put the label that you will use with \ref{} command in the braces of the \label{} command.
%
% \begin{figure}
% \end{figure}

% \begin{table}
% \end{tabular}
% \end{table}

\bibliography{CoPt_single_shot_ref}

\end{document}